\documentclass[final,3p,times,twocolumn]{elsarticle}
\usepackage{hhline}
\usepackage{epsfig}
\usepackage{graphicx}

\newcommand{\be}{\begin{equation}}
\newcommand{\ee}{\end{equation}}

\journal{Computational Physics Communications}
\begin{document}
\begin{frontmatter}

\title{A High-Performance Fortran Code to Calculate Spin- and Parity-Dependent 
Nuclear Level Densities}

\author[cmich]{R.A. Sen'kov}
\ead{senko1r@cmich.edu}

\author[cmich]{M. Horoi}

\author[msu]{and V.G. Zelevinsky}

\address[cmich]{Department of Physics, Central Michigan University, Mount Pleasant, Michigan 48859, USA}
\address[msu]{Department of Physics and Astronomy and National Superconducting Cyclotron Laboratory, Michigan State University, East Lansing, MI 48824-1321,USA}




\begin{abstract}
A high-performance Fortran code is developed to calculate the spin- and parity-dependent shell
model nuclear level densities.
The algorithm is based on the extension of methods of
statistical spectroscopy and implies exact
calculation of the first and second Hamiltonian moments for different
configurations at fixed spin and parity. The proton-neutron formalism is used.
We have applied the method for calculating the level densities for a set of nuclei
in the $sd$-, $pf$-, and $pf + g_{9/2}$ - model spaces. Examples of the calculations
for $^{28}$Si (in the $sd$-model space) and $^{64}$Ge (in the $pf + g_{9/2}$-model space)
are presented.
%
To illustrate the power of the method we estimate the ground state energy of $^{64}$Ge
in the larger model space $pf + g_{9/2}$, which is not accessible to direct shell model
diagonalization due to the prohibitively large dimension,
by comparing with the nuclear level densities at low excitation energy calculated in the smaller
model space $pf$.

\end{abstract}

\begin{keyword}
nuclear level density \sep parity \sep spin
\end{keyword}

\end{frontmatter}

\section{Introduction}

The nuclear reaction theory requires exact knowledge of nuclear level densities.
In the majority of cases, especially those important for reactions at stellar
conditions, experimental information on the excited nuclear states is not
sufficient and the calculations are typically using the estimates based on
the Hauser-Feshbach approach \cite{HF}, where the level density for specific
quantum numbers $J^{\pi}$ of nuclear spin and parity is a necessary ingredient.
The reaction rates can be very sensitive to the level density, especially in
the Gamow window of excitation energies around the particle threshold \cite{rauscher97,moller09}.

The theory of the nuclear level density has a long history starting with the combinatorial
calculation by Bethe \cite{bethe36} that leads to the the back-shifted Fermi gas
approximation \cite{cameron,ericson} that was improved over the years. The modern
approaches going essentially in the same direction 
[7-9]
use, instead of the Fermi-gas, an independent particle model in a mean field.
It is understood that many-body correlations may significantly change the resulting
level density, in particular moving down certain families of vibrational and rotational states.
At the same time, pure collective models used on top of the single-particle excitations
may suffer from double counting and have to be cut off in some way.

The many-body correlations are fully accounted for by the exact diagonalization of
the shell-model Hamiltonian. Here the problems come from the prohibitively large
dimensions that makes necessary to truncate the orbital space, a step not always
well controlled. Alternatively, one can use Monte-Carlo techniques 
[10-15], 
or other methods of statistical spectroscopy \cite{calvin-06-1,calvin-06-2}.
Most of these methods 
[2,10-12,18,19]
calculate the total {\sl density of states} and later use a spin-weight factor
that includes an energy-dependent cut-off parameter to extract the level density
for specific quantum numbers of spin, parity and isospin. Although there are recent
efforts to improve the accuracy of such parameterizations \cite{dorel-09}, it was
shown that the cut-off parameter has very large fluctuations at low excitation energy \cite{nic8den}.
The parity is usually taken as equally distributed, although there are attempts
\cite{alhassid00,mocelj-pm} to model the effect of the uneven parity-dependence of
the level densities at excitation energies of interest for nuclear astrophysics.

Recently, we developed a consistent approach 
[20,22-25]
to calculate the spin- and parity-dependent shell-model level
density. The new effective interactions for the appropriate model spaces
are developed starting with the $G$-matrix \cite{gmat} and fixing the
monopole terms or/and linear combinations of two-body matrix
elements to experimental data. Extending the efficient methods of
statistical spectroscopy \cite{french83,wongbook} we exactly
calculate the first and second moments of the Hamiltonian for different
configurations at fixed spin and parity. As a practical tool we use the exact
decomposition of many-body configurational space into classes corresponding
to different parity and number of harmonic oscillator excitations. An accurate
estimate of the shell-model ground state energy is required being generally
as time consuming as the previous steps. This stage can be improved by using the
exponential convergence method suggested and applied in
Refs. \cite{ecm,aecm1,aecm2}, or/and the recently developed projected configuration
interaction method \cite{pci1,pci2}. In reverse, some knowledge about the level density
can be helpful for extracting the ground state energy.

The code described in this paper is based on nuclear statistical spectroscopy \cite{wongbook}.
It allows one to calculate the spin- and parity-projected moments of the
nuclear shell-model Hamiltonian, which can be further used for
an accurate description of the level density up to about 15 MeV
excitation energy. It can be also applied to other mesoscopic systems,
such as interacting cold atoms in harmonic oscillator traps.
The code is parallelized using the Message Passing Interface (MPI) \cite{mpi}
and a master-slaves dynamical load-balancing approach. The parallel code was thoroughly tested on
the massively parallel computers at NERSC \cite{nersc}, and it shows very good scaling when
using up to 4000 cores.


The paper is organized as follows.
In Sec. II the method of fixed spin- and parity-dependent configuration moments is revisited.
The method allows to trace such quantum numbers as parity
and angular momentum explicitly. The extension of the algorithm to the proton-neutron formalism is
discussed in Sec. III.
In Sec. IV we introduce the structure of the program and supply the examples of input files.
Examples of calculations are presented and compared to exact shell model results in Sec. V.
Section VI is devoted to conclusions.

\section{Theory outline}

In this work we closely follow the approach proposed
in Ref \cite{our2010} (see also Refs. \cite{jtpden,jdenrc}).  For clarity we repeat here the main
ideas and equations we are going to use for calculating the level density.

According to the method of moments one can calculate
the density $\rho(E,\alpha)$ of levels with a given set of quantum numbers
as a function of excitation energy $E$ as a sum
\be \label{rho1}
\rho(E,\alpha) = \sum_\kappa D_{\alpha \kappa } \cdot G_{\alpha \kappa} (E),
\ee
where $\alpha = \{n, J, T_z, \pi \}$ includes all quantum numbers of interest,
namely the number of particles $n$ (protons and neutrons), total spin $J$,
isospin projection $T_{z}$, and parity $\pi$. The subscript
$\kappa$ represents a configuration of $n$ particles
distributed over $q$ spherical single-particle orbitals. Each
configuration $\kappa$ is fixed by a set of occupation
numbers, $ \kappa=\{\kappa_1, \kappa_2,... \; , \kappa_q \}, $
where $\kappa_j$ is the number of particles occupying the spherical single-particle
level $j$. The configuration has a certain number of particles, total isospin projection,
and parity. The sum in Eq. (\ref{rho1}) runs over all possible
configurations corresponding to given values of $n,T_z$, and $\pi$.
The dimension $D_{\alpha \kappa}$ equals the number of correctly antisymmetrized
many-fermion states with given $J$ that can be built for a given configuration $\kappa$.
The function $G_{\alpha \kappa}$
is a finite-range Gaussian defined as in \cite{jtpden}:
\begin{eqnarray}\label{frg}
G_{\alpha \kappa}(E)=G(E+E_{{\rm g.s.}}-E_{\alpha \kappa},\sigma_{\alpha \kappa}),\\
\label{frg2}
G(x,\sigma)= N \cdot \left\{
\begin{array}{ll}
\mbox{exp}\left( -x^2/2\sigma^2 \right) &, \; \; |x| \leq \eta \cdot \sigma \\
0 &, \; \; |x|> \eta \cdot \sigma \\
\end{array} \right. ,
\end{eqnarray}
where $E_{\alpha \kappa}$ and $\sigma_{\alpha \kappa}$ are the fixed-$J$ centroids
and widths, which will be defined later, $E_{{\rm g.s.}}$ is the ground state energy, $\eta$
is the cut-off parameter, and $N$ is the normalization factor corresponding to the
following condition: $\int_{-\infty}^{+\infty} G(x,\sigma) dx = 1$.

 A very important ingredient of the method is the accurate knowledge of
the ground state energy $E_{{\rm g.s.}}$. It is also necessary to find an optimal value of
the cut-off parameter $\eta$, see the discussion in \cite{our2010}.

Assuming a two-body shell-model Hamiltonian,
\be
\label{h}
H = \sum_i \epsilon_i a^\dag_i a_i + \frac{1}{4} \sum_{i j k l} V_{i j k l}
a^\dag_i a^\dag_j a_l a_k,
\ee
we have to calculate traces of the first and second power of this Hamiltonian,
Tr$[H]$ and Tr$[H^2]$, for each configuration $\kappa$ which will determine
the fixed-$J$ centroids and widths in Eq. (\ref{frg}):
\begin{eqnarray}
E_{\alpha \kappa} = \left< H \right>_{\alpha \kappa}, \\
\sigma_{\alpha \kappa} = \sqrt{ \left< H^2 \right>_{\alpha \kappa}
- \left< H \right>^2_{\alpha \kappa} },
\end{eqnarray}
where
\begin{eqnarray}
\label{tr1}
\left< H \right>_{\alpha \kappa} =
\mbox{Tr}^{(\alpha \kappa)}[H]/D_{\alpha \kappa},\\
\label{tr2}
\left< H^2 \right>_{\alpha \kappa} =
\mbox{Tr}^{(\alpha \kappa)}[H^2]/D_{\alpha \kappa}.
\end{eqnarray}
If the many-body states $\left|\nu, J \right>$ with a certain set of quantum numbers $\alpha\equiv\{\nu,J\}$, including spin $J$,
form a complete set for the configuration $\kappa$, the symbol of trace, $\mbox{Tr}^{(\alpha \kappa)}[\cdots]$,
means the sum of all diagonal matrix elements, $\sum \left< \nu, J | \cdots | \nu, J \right>$, within this subspace.
Technically, it is more convenient to derive these traces in a basis with a fixed
spin projection $\left| \nu, M_z \right>$, $\mbox{Tr}^{(M_z)}[\cdots]$, rather than in the basis with
fixed total spin $\left| \nu, J \right>$, $\mbox{Tr}^{(J)}[\cdots]$. $J$-traces can be easily expressed
through the $M_z$-traces, given the rotational symmetry of the Hamiltonian,
\be
\label{mztoJ}
\mbox{Tr}^{(J)} [\cdots]
= \mbox{Tr}^{(M_z)} [\cdots]{\Big \vert}_{{}_{M_z=J}} -
\mbox{Tr}^{(M_z)} [\cdots] {\Big \vert}_{{}_{M_z=J+1}}.
\ee
For simplicity, in Eq. (\ref{mztoJ}) we omitted all quantum numbers, except the projection $M_z$ and
the total spin $J$.

\begin{figure}
\begin{center}
\includegraphics[width=0.48\textwidth]{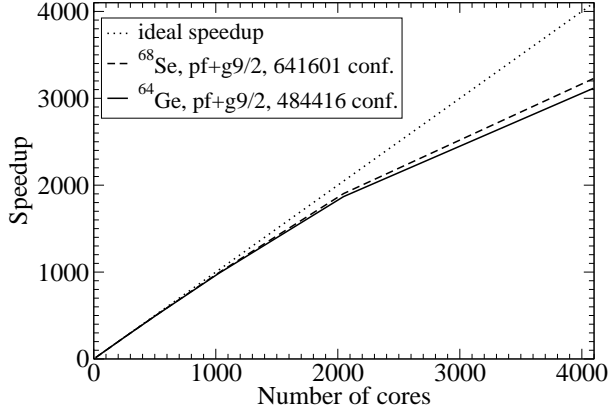}
\caption{Speedup is defined as $T_1/T_n$, where $T_n$ is the calculation time,
when $n$ processors were used. These calculations were performed on the FRANKLIN supercomputer at the National
Energy Research Scientific Computing Center (NERSC) \cite{nersc}.}\label{spup}
\end{center}
\end{figure}

Hereafter we use the label $\alpha$ to denote a set of quantum numbers that includes
either the
fixed $M_z$ or the fixed $J$, keeping in mind that Eq. (\ref{mztoJ})
can always connect them.
In every important case we will point out which set of quantum numbers is used.
Following the approach of \cite{jaq79}, we can obtain the following
expressions for the traces in Eqs. (\ref{tr1}) and (\ref{tr2}):
\begin{eqnarray}
\label{tr1c}
\mbox{Tr}^{(\alpha \kappa)}[H] = \sum_i \epsilon_i  D^{[i]}_{\alpha \kappa}
+ \sum_{i<j} V_{ijij}D^{[ij]}_{\alpha \kappa},\;\;\;\;\\
\nonumber
\mbox{Tr}^{(\alpha \kappa)}[H^2] = \sum_i \epsilon^2_i  D^{[i]}_{\alpha \kappa} + \\
\nonumber
+ \sum_{i<j} \left[ 2 \epsilon_i \epsilon_j +  2 (\epsilon_i+\epsilon_j) V_{ijij} + \sum_{q<l}
V^2_{ijql} \right] D^{[ij]}_{\alpha \kappa} + \\
\nonumber
+\sum_{(i<l)\ne l} \left[ \sum_q \left( 2V_{liiq}V_{ljjq}-V^2_{ijql}\right)
+ 2 \epsilon_l V_{ijij}\right] D^{[ijl]}_{\alpha \kappa} + \\
\label{tr2c}
+ \sum_{(i<j)\ne(q<l)} \left[ V^2_{ijql}+V_{ijij}V_{qlql}-4V_{qiil}V_{qjjl}\right]
D^{[ijql]}_{\alpha \kappa},\;\;\;\;
\end{eqnarray}
where $i,j,l,$ and $q$ are single-particle states with certain
spin projections and possible occupation numbers equal 0 or 1.

Notice that the single-particle orbitals we have used  to define
the configurations in Eq. (\ref{rho1}) can host all particles with all possible
spin projections corresponding to spin of the orbital. The dimension factor
$D^{[i]}_{\alpha \kappa} = \mbox{Tr}^{(\alpha \kappa)}[a^\dag_i a_i]$
can be interpreted as a number of many-body states, possible for the configuration $\kappa$,
with the fixed projection $M_z$ (if we consider $M_z$-traces) and under the condition that
the single-particle state $i$ is occupied:
$$
D^{[ij]}_{\alpha \kappa}= \mbox{Tr}^{(\alpha \kappa)}[a^\dag_i a^\dag_j a_j a_i],
$$
\begin{equation}
D^{[ijq]}_{\alpha \kappa} = \mbox{Tr}^{(\alpha \kappa)}[a^\dag_i a^\dag_j a^\dag_q a_q a_j a_i], ...
\end{equation}
These $D$-structures were called {\sl propagation functions} in \cite{jaq79}.
For completeness, we repeat here the recipe used for calculating them.
One can show \cite{jaq79} that
\be
\label{rec}
D^{[r_1 r_2 \cdots r_s]}_{\alpha \kappa} = \sum_{s \leq t\leq n}(-1)^{t-s}\sum_{t_1+\cdots +t_s=t}
D_{\alpha' \kappa'},
\ee
where all $t_i$ are non-negative integers, the configuration $\kappa'=\{\kappa'_1, \kappa'_2,... \; , \kappa'_q \}$
can be derived from the original configuration $\kappa=\{\kappa_1, \kappa_2,... \; , \kappa_q \}$
by removing $t$ particles from the single-particle states $r_1, r_2, \cdots r_s$.
A formal expression for the new configuration $\kappa'$ can be written as follows:
\be
\kappa'_j=\kappa_j - \sum_{i \; (r_i \in j)} t_i,
\ee
where the sum includes only
those values of $i$ for which the corresponding single-particle state $r_i$ belongs
to the single-particle level $j$.
We also assume that all the occupation numbers $\kappa'_j$ must be positive, which imposes certain
restrictions on the possible values of the amplitudes $t_i$. For every new configuration $\kappa'$ one
can easily define new quantum numbers, $\alpha'=\{n' M'_z T'_z \pi' \}$, entering Eq. (\ref{rec}).
Examples are the new number of particles $n'=n-t$ and the new spin projection,
\be
M'_z=M_z-t_1 m_{r_1}-t_2 m_{r_2}-\cdots-t_s m_{r_s},
\ee
where $m_{r_i}$ is the $M_z$ projection of the single-particle state $r_i$.
The new isospin $T'_z$ and parity $\pi'$ are defined similarly.

\section{The algorithm for the method of moments in the proton-neutron formalism}

Here we describe some technical features of the algorithm developed for the calculation of
the nuclear level density. We treat protons and neutrons separately, so that
 the basis of many-body wave functions is represented by a product of proton and neutron
parts:
\be
\label{bwf}
|\nu, M_z\rangle = |\nu_p, M^{(p)}_{z}\rangle \cdot |\nu_n, M^{(n)}_{z}\rangle,
\ee
where $M^{(p)}_{z}+M^{(n)}_{z}=M_z$. The wave functions (\ref{bwf}) have the
fixed isospin projection $T_z$, but do not have a certain isospin $T$. As
we already mentioned, it is more convenient to use the basis of the wave functions with the fixed
spin projection $M_z$, rather than the one with the fixed spin $J$.

One could gain essential advantages from such a separation of the basis.
One of them is related to the number of configurations that appear
in the sum of Eq. (\ref{rho1}). Naturally, the number of configurations with fixed
$T_z$ is much greater than the number of configurations with fixed isospin. This
allows the use of many-cores computers with greater efficiency.
In other words, the calculation of the sum in Eq. (\ref{rho1}) with a larger number of
configurations
can be more efficiently distributed over a larger number of processors. Fig. \ref{spup}
presents the speedup (calculation speed gain) as a function of the number of used processors.
One can see that the case with the larger number of configurations, ${}^{68}$Se, scales better
than the case with the lower number of configurations, ${}^{64}$Ge. Up to 2000 cores, the
speedup is almost perfect (the dotted line presents an ideal speedup). At this point the
calculation time is about 1-2 minutes and further improvement is
hardly achievable.

Another significant advantage of the proton-neutron formalism is the new algorithm of calculating the dimensions
$D_{\alpha \kappa}$, $D^{[i]}_{\alpha \kappa}$, $D^{[i j]}_{\alpha \kappa}$, etc.
Because of the proton-neutron separation one can calculate all proton and neutron dimensions separately.
Later, the dimensions we are interested in can be easily constructed
from the proton and neutron parts using the convolution,
\be
\label{drel}
D_{M_z \kappa} = \sum_{M^{(p)}_z+M^{(n)}_z=M_z} D_{M^{(p)}_z \kappa_p} \cdot D_{M^{(n)}_z \kappa_n},
\ee
where, instead of the whole set of quantum numbers $\alpha$, only the spin projection $M_z$ was explicitly indicated.
Here $\kappa_p$ and $\kappa_n$ are the proton and neutron parts of the configuration $\kappa$.
Eq. (\ref{drel}) can be easily applied to all types of dimensions, $D^{[\cdots]}_{\alpha \ldots}$,
needed in the formalism of section II. The advantage comes
from the fact that one can calculate and keep in memory all proton and neutron dimensions,
$D_{M^{(p)}_z \kappa_p}$ and $D_{M^{(n)}_z \kappa_n}$, for all possible projections $M^{(p)}_z$ and $M^{(n)}_z$,
and for all possible configurations $\kappa_p$ and $\kappa_n$.
Afterwards, using Eqs. (\ref{drel}) and (\ref{rec}), one can calculate very fast all the dimensions:
$D_{\alpha \kappa}$, $D^{[i]}_{\alpha \kappa}$, $D^{[i j]}_{\alpha \kappa}$, etc... , for all $M_z$ and $J$.

One more technical detail, which allows a significant speed up of the algorithm, is  that by using the proton-neutron separation one can avoid  multiple computations of the most time consuming structures, such as $D^{[i j q l]}_{\alpha \kappa}$. Let us consider a case when all four single-particle states $\{i j q l \}$ are protons. One can then use an equation similar to Eq. (\ref{drel}),
\be
\label{ndrel}
D^{[i j q l]}_{M_z \kappa} = \sum_{M^{(p)}_z+M^{(n)}_z=M_z} D^{[i j q l]}_{M^{(p)}_z \kappa_p} \cdot D_{M^{(n)}_z \kappa_n}.
\ee
For all configurations $\kappa$ that have the same proton parts $\kappa_p$ one would have to recalculate $D^{[i j q l]}_{M^{(p)}_z \kappa_p}$  for each neutron configuration. Alternatively, one can calculate $D^{[i j q l]}_{M^{(p)}_z \kappa_p}$ only once, and store the results in memory. That strategy, however, would require a large amount of storage. More efficiently, one can only store the contributions of the  $D^{[i j q l]}_{\alpha \kappa}$ structures to the width,  Eq. (\ref{tr2c}), that is, one can only store  the following $T$-structures,
$$
T_{M^{(p)}_z \kappa_p}=\sum_{(i<j)\ne(q<l)} \left[ V^2_{ijql}+V_{ijij}V_{qlql} \right.
$$
\be \label{prc6}
\left. - 4V_{qiil}V_{qjjl}\right] D^{[ijql]}_{M^{(p)}_z \kappa_p},
\ee
where all single-particle states are protons. Thus, instead of using Eq. (\ref{ndrel}) one can calculate the contribution to the width directly via the convolution,
\be\label{tr2new}
\mbox{Tr}^{(\alpha \kappa)}[H^2] = \dots + \sum_{M^{(p)}_z+M^{(n)}_z=M_z} T_{M^{(p)}_z \kappa_p} \cdot D_{M^{(n)}_z \kappa_n},
\ee
which is very similar to Eqs. (\ref{drel}) and (\ref{ndrel}).
The new approach avoids multiple calculations of $D^{[i j q l]}_{M^{(p)}_z \kappa_p}$. Storing the structures
Eq. (\ref{prc6}), one may significantly speed up the algorithm for large cases, such as ${}^{68}$Se in $pf+g_{9/2}$ model space. The downside is that the calculation of the $T$-structures, $T_{M^{(p)}_z \kappa_p}, T_{M^{(n)}_z \kappa_n}$, does not always scale well on a large number of cores, since the number of these $T$-structures is much smaller than the total number of configurations.

%
%
%

%
%
%
%
%

\section{Description of the program}

The program consists of two separate codes. The first code
is called MM, which is the main code in the program. MM code
performs calculation of the first and second moments for all
the configurations within the given range of spins and for certain parity.
It is the most complicated and resource demanding part of the program.
MM requires parallel computing.
The second code is very simple and fast. It takes the output of the
MM code (the first and second moments) and builds the nuclear
level  densities according to Eqs. (\ref{rho1}-\ref{frg2}). It does not
require parallel computing.

Below we will concentrate only on description of the MM code. The detailed
instructions on how
to use the second code can be found in the {\bf readme.txt} file, which is
in the main project directory.

\subsection{The structure of the MM code}

In the MM code, the calculation of the first and second moments for a given
nucleus is carried out. To compile this code simply follow the instructions
in the {\bf readme.txt} file (or type {\it make}) and then run the executable 
file {\bf mm.out}.

\begin{figure}
\begin{center}
\includegraphics[width=0.45\textwidth]{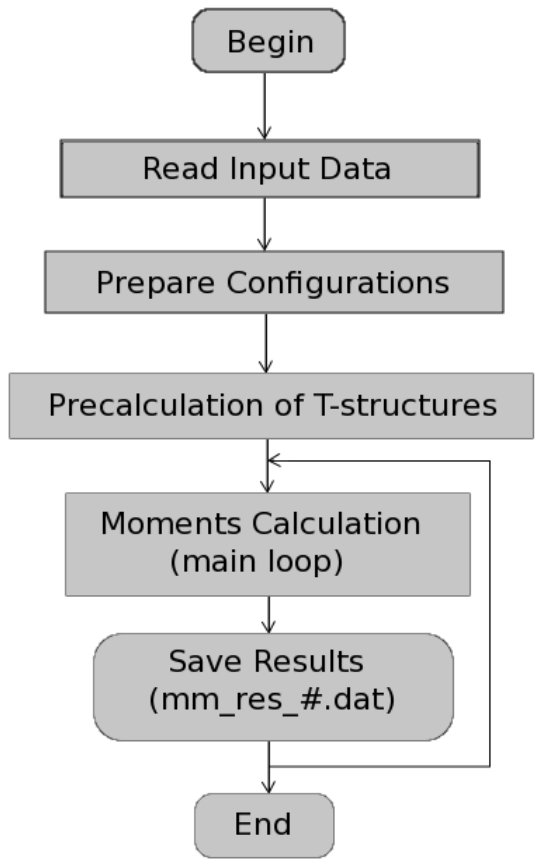}
\caption{ The flowchart of the program. See text for details. }\label{flowchart}
\end{center}
\end{figure}

The code contains four files: {\bf mm\_1.15.f90} (1.15 is the current version)
contains the main subroutines and calls subroutines from other files to perform
the calculation; {\bf interaction.f90} reads the interaction file; {\bf angularme.for} contains
subroutines to work with the angular matrix elements,
for example, to calculate Clebsch-Gordan coefficients; {\bf qsort.f} contains
quick-sort subroutines.

 The flowchart of the {\bf mm\_1.15.f90} is shown in Fig.\ref{flowchart}, and
 some important subroutines in the {\bf mm\_1.15.f90} are listed below.

 The subroutine {\bf prepare\_interaction} calculates the $T$-structures according to Eq. (\ref{prc6}). This is an important part of the code which allows to speed up the program significantly.

 The subroutine {\bf cc\_density\_calc} contains the main loop over the configurations (the loop back in Fig. 2),
 calculates the first and second moments based on Eqs. (\ref{tr1},\ref{tr2},\ref{tr1c},\ref{tr2new}),
 and saves the results. This subroutine
 uses the $T$-structures precalculated in the subroutine {\bf prepare\_interaction}.

 Both subroutines require parallel computing. The simplest ``Master-Slave" parallel programming paradigm with Dynamic Load Balancing was used.

\subsection{MM input}

The input files include {\bf input.dat} and the files that define the single-particle model
 space as well as the interaction in this model space. A typical {\bf input.dat} file that 
 specifies the parameters in the code is listed below. The parameters are followed by their meaning:

\bigskip

\begin{tabular}{ll}
6 6 & ! $Z\; N$ \\
1 0 22 2 & ! $P\; 2J_{{\rm min}}\; 2J_{{\rm max}}\; 2J_{{\rm step}}$\\
int/sd.spl & ! single-particle model space\\
int/usd.int & ! interaction
\end{tabular}

\bigskip

Here $Z$ and $N$ are the number of protons and neutrons, respectively, in the valence space;
$P$ is parity ($P=1$ corresponds to positive parity and
$P=-1$ to negative parity); $2J_{{\rm min}}$, $2J_{{\rm max}}$, and $2J_{{\rm step}}$ define the range of
total spin for which the moments are to be calculated: the spin changes from the
minimum value $J_{{\rm min}}$ to the maximum value $J_{{\rm max}}$ with the step $J_{{\rm step}}$.
In the example listed above the total spin changes from 0 to 11 with the step 1.
The single-particle model space and the interaction are defined by two
separate files and {\bf input.dat} must have the names of these two files
similar to the shown example. Detailed description of {\bf *.spl} and {\bf *.int} files
is given  in the  {\bf readme.txt} file.

The above example describes the $^{28}$Si nucleus in the $sd-$model space
with the USD interaction. The moments will be calculated for positive parity and
for all possible spins from 0 to 11.

\subsection{MM output}

The main output of the MM code is presented in files {\bf mm\_res\_\#.dat}.
These files are enumerated by spin number \#, for example {\bf mm\_res\_0.dat} corresponds
to spin $J=0$, {\bf mm\_res\_1.dat} corresponds to spin $J=1/2$, and so on (for more details
see the {\bf readme.txt} file). Each output file contains the data needed for the
density calculation: dimensions, first and second moments.

Another output file {\bf mm\_conf.dat} contains information about the configurations.
This information is not used for the density calculation, but could be useful for checking and testing
purposes.

\subsection{Density calculation}

After all the moments are prepared, the density can be calculated with the code
{\bf den.out}. The full description of density calculation can be found in the
{\bf readme.txt} file. We just repeat once more that there are several external
parameters that need to be prepared before the calculation. These
parameters cannot be defined within the method of moments, namely:
$\eta$ - the cut-off parameter, $E_{{\rm g.s.}}$ - the ground state energy, and the
energy interval $(E_{{\rm min}}$, $E_{{\rm max}})$ for calculating the
level density. For more details see the {\bf readme.txt} file.

\section{Examples}

\begin{table}[ht]
\begin{center}\label{ttime}

\begin{tabular}{||c|c|c|c||}
\hhline{|t:=:=:=:=:t|}
\hhline{||---||}
\rule{0cm}{0.33cm} Element & Space & Total dim & Elapsed time (sec)\\
\hhline{||----||}
\rule{0cm}{0.33cm} ${}^{70}$Br & $pf+g_{9/2}$ & $10^{15}$ & $1.07 \cdot 10^4$\\
\hhline{||----||}
\rule{0cm}{0.33cm} ${}^{68}$Se & $pf+g_{9/2}$ & $10^{15}$  & $1.03 \cdot 10^4$\\
\hhline{||----||}
\rule{0cm}{0.33cm} ${}^{64}$Ge & $pf+g_{9/2}$ & $10^{14}$ & $0.76 \cdot 10^4$\\
\hhline{||----||}
\rule{0cm}{0.33cm} ${}^{60}$Zn & $pf$ & $10^{11}$ & 37.4\\
\hhline{||----||}
\rule{0cm}{0.33cm} ${}^{52}$Fe & $pf$ & $10^{10}$ & 13.6\\
\hhline{||----||}
\rule{0cm}{0.33cm} ${}^{28}$Si & $sd$ & $10^{6}$ &0.7\\
\hhline{|b:=:=:=:=:b|}
\end{tabular}

\caption{Elapsed times of nuclear level density calculations
(for all $J$, positive parity) with the moments method code. The calculations
were done on a 16 cores machine with 2.8 GHz CPU frequency.}\label{tb1}

\end{center}
\end{table}

Table \ref{tb1} presents calculation times for different nuclei calculated in different shell-model spaces.
The calculations were done on a 16 cores machine with 2.8 GHz CPU frequency. One core ('master')
distributed all the work between other 15 cores ('slaves'). One can emphasize here that the listed times
correspond to calculations of the nuclear level densities for all $J$ and for positive parity.
For the case of ${}^{68}$Se the largest $m$-scheme dimension is about $10^{15}$.  For each $J$ the $m$-scheme
dimensions vary from $10^{12}$ to $10^{14}$, which makes direct
diagonalization impossible. Using the moments method and our algorithm we are able to calculate
the shapes of nuclear densities for ${}^{68}$Se in less then three hours on a 16 cores machine.
For a number of processors reaching one thousand, it will take only few minutes to complete the calculation.

\subsection{$^{28}$Si, $sd-$model space}

\begin{figure*}
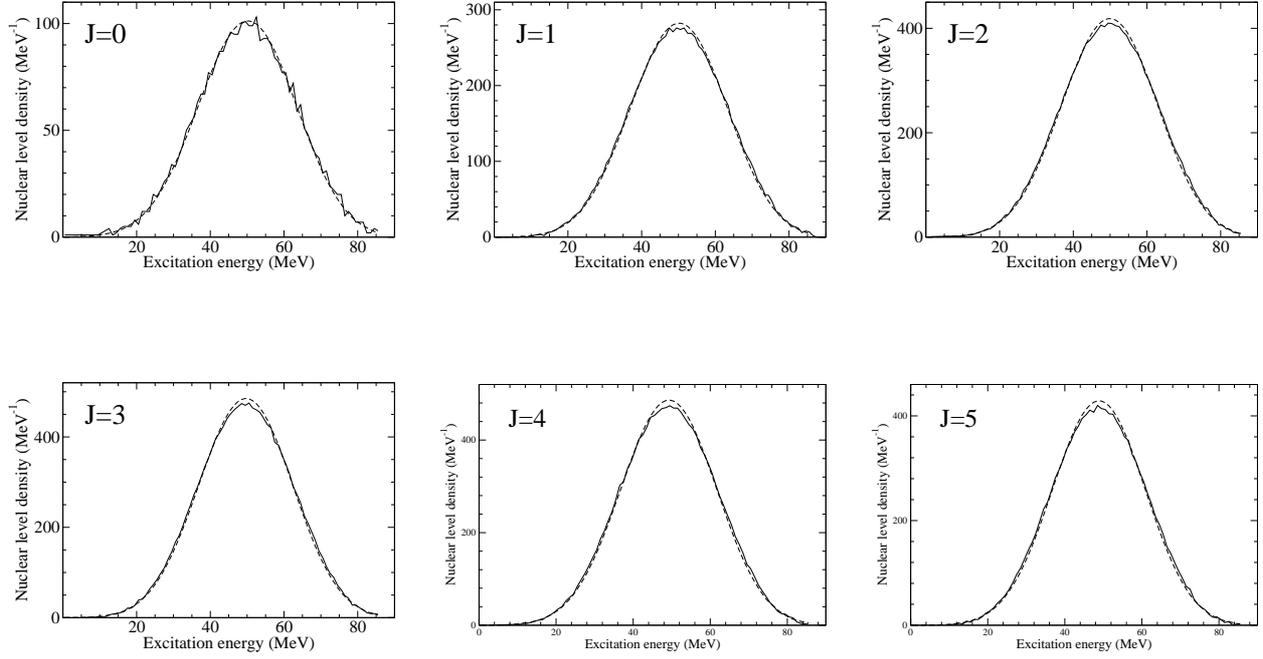

\includegraphics[width=0.31\textwidth]{si0.eps}
\hfill
\includegraphics[width=0.31\textwidth]{si2.eps}
\hfill
\includegraphics[width=0.31\textwidth]{si4.eps}
\\

\vspace{1.0cm}

\includegraphics[width=0.31\textwidth]{si6.eps}
\hfill
\includegraphics[width=0.31\textwidth]{si8.eps}
\hfill
\includegraphics[width=0.31\textwidth]{si10.eps}
\\

%

\caption{${}^{28}$Si, parity=+1. Comparison of nuclear level densities
between the exact shell model (solid line) and the moments method (dashed line).
Cut-off parameter $\eta=2.8$, interaction: USD, $sd$-shell. }\label{si28}

\end{figure*}

As a first example
we consider the level density for the  ${}^{28}$Si nucleus in the $sd$-shell model
space, where we use the USD interaction \cite{usd}. Fig. \ref{si28} presents
the comparison of the exact shell-model level densities for different spins (solid lines) with
those obtained with the moments methods (dashed lines).

Eqs. (\ref{rho1}) and (\ref{frg}) require
the knowledge of the ground state energy $E_{{\rm g.s.}}$ and the cut-off parameter $\eta$.
While the ground state energy of ${}^{28}$Si can be calculated in this case using
the standard shell model, $E_{{\rm g.s.}}=-135.94$ MeV, for the value of the cut-off parameter $\eta$ we have
only a general idea that it should be around 3 \cite{jtpden,jdenrc}.
For a better description of level densities in the moments method
we can adjust the $\eta$ parameter to optimally reproduce the exact shell-model densities.
From Fig. \ref{si28} one can see that choosing $\eta=2.8$,
the level densities of the moments method reproduce quite well the exact shell-model level densities.
The cut-off parameter plays a role similar to that of the width in a Gaussian distribution. Indeed, if we increase
the cut-off parameter, the density becomes wider and lower, while decreasing it
leads to a narrowing of the density.
One should also mention that the exact
spin- and parity-dependent shell-model densities were calculated with the NuShellX code \cite{nushellx}.



\subsection{$^{64}$Ge, $pf-$ and $pf+g9/2-$model spaces}

\begin{figure*}
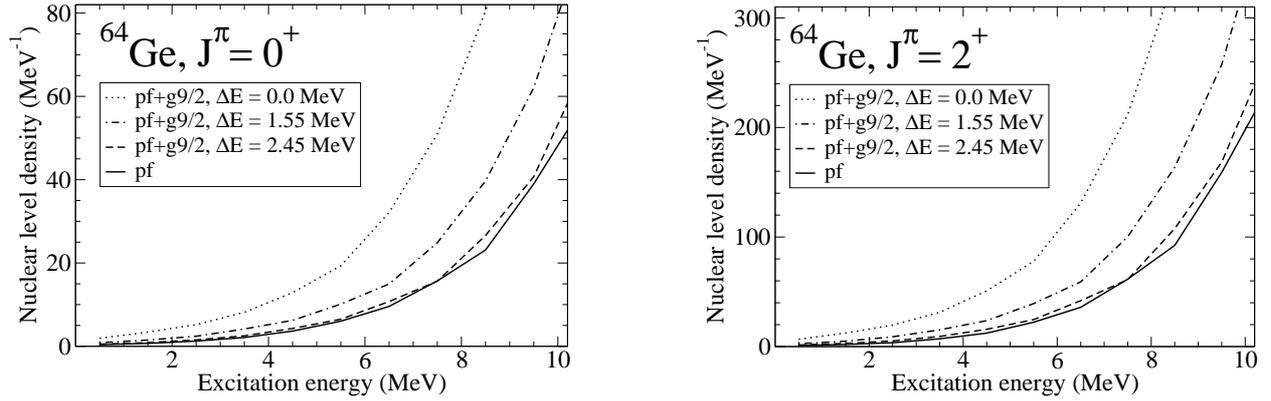

\includegraphics[width=0.45\textwidth]{ge0.eps}
\hfill
\includegraphics[width=0.45\textwidth]{ge4.eps}
\\
\caption{Level densities for the $^{64}$Ge nucleus, $J=0,2$ and positive parity. Solid line presents the calculation in the $pf$-shell with the GXPF1A interaction. For this calculation we know the ground state
energy, $E_{{\rm g.s.}}(pf)=-304.25$ MeV. Other three lines present calculations in the large model space, when
level $g_{9/2}$ is added. The ground state energy for these cases is $E_{{\rm g.s.}}(pf+g_{9/2})=E_{gs}(pf)-\Delta E$,
where $\Delta E$ is the energy shift. The cut-off parameter is $\eta=2.6$.}\label{ge64}
\end{figure*}

As mentioned in the Introduction, one could envision using information
from the level densities to extract with a good approximation the ground state energies.
Using our algorithm and
the moments method one can easily calculate the nuclear level density
for any nucleus that can be described in the $pf+g_{9/2}$ model space.
The Hamiltonian used for this
model space was built starting with the GXPF1A interaction for the $pf$ model space, to which
the $G$-matrix elements that describe the interaction between the $pf$ orbits and $g_{9/2}$ orbit
were added. The single-particle energy for the $g_{9/2}$ orbit was fixed at $-$0.637 MeV.
Even for the worst case, the calculation takes about three hours for sixteen processors and only
few minutes for one thousand processors.
Fig. \ref{ge64} presents the results obtained for ${}^{64}$Ge, nucleus that is
believed to be a ``waiting-point" along the $rp$-process path 
[39-41]. 
We only present the densities for $J=0,2$ and positive parity.

The corresponding {\bf input.dat} file for $^{64}$Ge in $pf-$model space looks like

\bigskip

\begin{tabular}{ll}
12 12 & ! $Z\; N$ \\
1 0 22 2 & ! $P\; 2J_{{\rm min}}\; 2J_{{\rm max}}\; 2J_{{\rm step}}$\\
int/pf.spl & ! single-particle model space\\
int/gx1a.int & ! interaction
\end{tabular}

\bigskip

Here we have twelve protons and twelve neutrons in the $pf$-model space. The calculation is done
for all spins between $J=0$ and $J=11$ and positive parity. The single-particle space is defined
in the {\bf pf.spl} file and the interaction is in {\bf gx1a.int}. For the $pf+g9/2$ model space we need to
change the single-particle file to {\bf pfg9.spl} and the interaction file to {\bf pfg9.int}.

It is important to notice that in the $pf$ model space the shell-model calculations of the ground state
energies can be done. For ${}^{64}$Ge in the $pf$-shell we obtain the following ground state energy:
\be
E_{{\rm g.s.}}(pf)=-304.25 \mbox{ MeV.}
\ee
Using this ground state energy and
 the cut-off parameter $\eta=2.6$, we are able to calculate the level densities
according to Eqs. (\ref{rho1}) and (\ref{frg}). The solid lines in Fig. \ref{ge64}
represent the density in the $pf$-shell.

To calculate the same level density in the $pf+g_{9/2}$ model space we have
to adjust the ground state energy and the cut-off parameter for this space.
For the cut-off parameter we use the same value, $\eta=2.6$, but it is practically impossible
to calculate by shell-model diagonalization the ground state energy since the dimension is too large.
The ground state energy for the larger
model space, that is $pf+g_{9/2}$, must be lower compared to the ground state energy for the smaller model space,
that is $pf$. Let us introduce this energy difference, $\Delta E$, as
\be
E_{{\rm g.s.}}(pf+q_{9/2})=E_{{\rm g.s.}}(pf)-\Delta E.
\ee
The dotted lines in Fig. \ref{ge64} show the level
densities if we keep the ground state energy for $pf+g_{9/2}$ model space as it was in the $pf$ case
keeping $\Delta E = 0$.
It is natural to expect only small differences between the level densities calculated in those
two model spaces at low excitation energy since in the $pf+g_{9/2}$ model space we use the same GXPF1A interaction
for the $pf$ subspace.
By decreasing the ground state energies for the $pf+g_{9/2}$ model space (introducing non-zero $\Delta E$),
one gets the dashed lines on Fig. \ref{ge64}. The dash-dotted lines there correspond to
ground state energy $E_{{\rm g.s.}}(pf+q_{9/2})=-305.8$ MeV of ${}^{64}$Ge, which was obtained by a truncated shell model calculation with up to 6 particles excited from the $f_{7/2}$ orbits and/or into the $g_{9/2}$ orbit. The
$m$-scheme dimension in this calculation, $13.5\times10^{9}$, is at the upper limit of the state of the
art shell-model calculation. As one can see, this value does not describe satisfactory the level densities at
low excitation energy.  In order to make the low-lying part of the two densities very close
(dashed and solid lines on Fig. \ref{ge64}), one has to adjust the
ground state energy for the $pf+g_{9/2}$ model space to the following value:
\be
\label{enge}
E_{{\rm g.s.}}(pf+g_{9/2})=-306.7 \; \mbox{ MeV for } \; {}^{64} \; \mbox{Ge}, \\
\ee
The ``low-lying part of the density" should be chosen such that the excitations
to the $g_{9/2}$ orbit do not give a significant contribution. For these cases we use
the interval 3-6 MeV in excitation energy.
We conclude that the adjustment of Eq. (\ref{enge}) can be treated
as a method for estimating the ground state energies in larger spaces; for more details see \cite{our2010}.

\section{Summary}
In summary, we have developed an efficient Fortran code for calculating the centroids and widths
of the shell-model spin- and parity-dependent configurations,
which can be used for calculating the nuclear level densities.
The code is parallelized using the Message Passing Interface (MPI) \cite{mpi}
and a master-slaves dynamical load-balancing approach. The parallel code was thoroughly tested on
the massively parallel computers at NERSC \cite{nersc}, and it shows very good scaling when
using up to 4000 cores.
The algorithm used takes advantage of the separation of the model space in
neutron and proton subspaces. This separation provides two important advantages:
(i) the exponentially exploding dimensions and propagators can be calculated more
efficiently in proton and neutron subspaces, and the full results can
be recovered via simple convolutions;
(ii) the number of configurations is significantly increased in the
proton-neutron formalism, considerably improving the scalability of the algorithm on
massively parallel computers. Our tests indicate almost perfect scaling for up to
4000 cores.
The new algorithm is so fast that the bottleneck of the calculation is now that of the ground state
energy. That is why we could not test our algorithm for cases that take more than one minute on
4000 cores.
Therefore, we investigated the possibility of using the calculated shapes of the nuclear level
densities to extract the ground state energy. We showed that by incrementing the model space 
and the effective interaction,
and imposing the condition that the level density does not change at low expectation energy,
one can reliably predict the ground state energy, and further the full level density. This new method
of extracting the shell model ground state energy for model spaces whose dimensions
are unmanageable for direct diagonalization opens new opportunities for calculating shell
model level densities of heavier nuclei of interest for nuclear astrophysics,
nuclear energy and medical physics applications.

A further development of the application of statistical spectroscopy to nuclear level density
is the removal of center-of-mass spurious states from the level density for shell model spaces
that allow complete factorization of the center-of-mass and intrinsic wave functions.
A new algorithm implementing this idea was recently presented \cite{shzplb}, and a high
performance code was developed. This new code could be made available upon request.

\section{Acknowledgments}

R.A.S. and M.H. would like to acknowledge the DOE UNEDF grant DE-FC02-09ER41584 for support.
M.H. and V.Z. acknowledge support from the NSF grant PHY-1068217.

\end{document}